\documentclass[11pt]{article}

\usepackage{amsmath}
\usepackage{amsfonts}
\usepackage{amssymb}
\usepackage{graphicx}
\usepackage{supertabular}
\usepackage{setspace}
\usepackage{xcolor}
\usepackage{array}
\usepackage{amsthm}
\usepackage{bm}
\usepackage{latexsym}
\usepackage{mathptmx}
\usepackage{hyperref}
\hypersetup{colorlinks=false}

\RequirePackage{cite}%
\renewcommand{\citeleft}{\bgroup\normalfont[}%
\renewcommand{\citeright}{]\egroup}%

\newcommand{\nin}{\noindent}

\newcommand{\be}{\begin{equation}}
\newcommand{\ee}{\end{equation}}
\newcommand{\ba}{\begin{eqnarray}}
\newcommand{\ea}{\end{eqnarray}}
\newcommand{\bal}{\begin{align}}
\newcommand{\eal}{\end{align}}

\newcommand{\dd}{{\rm d}}

\newcommand{\al}{\alpha}

\newcommand{\La}{\Lambda}
\newcommand{\bt}{\beta}

\newcommand{\ro}{\rho}
\newcommand{\ep}{\epsilon}

\newcommand{\ta}{\theta}

\newcommand{\Si}{\Sigma}
\newcommand{\De}{\Delta}

\newcommand{\Om}{\Omega}

\newcommand{\bw}{\begin{widetext}}
\newcommand{\ew}{\end{widetext}}

\def\abh{black hole }
\def\bh{black holes }

\def\BH{black holes}

\def\dS{de Sitter }
\def\S{Schwarzschild }

\begin{document}


\title{{\textbf{Regular and conformal regular cores for static and rotating solutions}}}

\author{Mustapha Azreg-A\"{\i}nou
\\Ba\c{s}kent University, Department of Mathematics, Ba\u{g}l\i ca Campus, Ankara, Turkey}
\date{}

\maketitle

\begin{abstract}
Using a new metric for generating rotating solutions, we derive in a general fashion the solution of an imperfect fluid and that of its conformal homolog. We discuss the conditions that the stress-energy tensors and invariant scalars be regular. On classical physical grounds, it is stressed that, conformal fluids used as cores for static or rotating solutions, are exempt from any malicious behavior in that they are finite and defined everywhere.

\vspace{3mm}

\nin {\footnotesize\textbf{PACS numbers:} 04.70.Bw, 04.20.-q, 97.60.Lf, 02.30.Jr}

\vspace{-3mm} \nin \line(1,0){430} 
\end{abstract}

\section{Introduction\label{sec1}}

The quest for rotating solutions has always been a fastidious task. It took more than two decades to discover the rotating solution of Van Stockum~\cite{Van} and more than forty years to derive that of Kerr~\cite{Kerr} since the foundation of General Relativity in 1916. Several partial methods have been put forward to construct rotating solutions~\cite{Van}-\cite{phantom1} but no general method seems to be available. This work is no exception and presents a novel partial method for generating rotating solutions from static ones. However, the method will allow us (1) to generate rotating solutions without appealing to linear approximations~\cite{linear} and (2) to apply the matching methods~\cite{IrinaD,Magli,Lemos} to regular \abh cores as well as to wormhole cores~\cite{phantom1,regular,phantom2}. The excellent paper by Lemos and Zanchin offers an up-to-date classification of the existing matching methods, discusses the types of regular \bh derived so far and presents new electrically charged solutions with a regular de Sitter core~\cite{Lemos}. The present method reduces the task of finding a rotating solution to that of finding a two-variable function that is a solution to two second order partial differential equations.

We work with $R^{\mu}{}_{\nu\rho\sigma}= -\partial_{\sigma}\Gamma^{\mu}{}_{\nu\rho}+\cdots$ ($\mu=1\to4$)
and a metric $g_{\mu\nu}$ with signature ($+,-,-,-$). We make all necessary conventions such that the field equations take the form $G_{\mu\nu}=T_{\mu\nu}$.

We consider a fluid without heat flux, the stress-energy tensor (SET) of which admits the decomposition
\begin{equation}\label{1}
    T^{\mu\nu}=\ep u^{\mu}u^{\nu}+p_2e^{\mu}_2e^{\nu}_2+p_3e^{\mu}_3e^{\nu}_3+p_4e^{\mu}_4e^{\nu}_4
\end{equation}
where $\ep$ is the mass density and ($p_1,\,p_2,\,p_3$) are the components of the pressure. We have preferred the notation $u^{\mu}$, instead of $e^{\mu}_1$, which is the four-velocity of the fluid. The four-vectors are mutually perpendicular and normalized: $u^{\mu}u_{\mu}=1$, $e^{\mu}_ie_{i\mu}=-1$ ($i=2\to4$). If the fluid is perfect, $p_2=p_3=p_4\equiv p$, then the completeness relation, $g^{\mu\nu}=u^{\mu}u^{\nu}-(e^{\mu}_2e^{\nu}_2+e^{\mu}_3e^{\nu}_3+e^{\mu}_4e^{\nu}_4)$, leads to $T^{\mu\nu}=(\ep +p) u^{\mu}u^{\nu}-pg^{\mu\nu}$.

Given a static spherically symmetric solution to the field equations in spherical coordinates:
\begin{equation}\label{3}
    \dd s^2 = G(r)\dd t^2 - \frac{\dd r^2}{F(r)} - H(r)(\dd \ta^2+\sin^2\ta\dd \phi^2)
\end{equation}
we generate a stationary rotating solution, the metric of which, written in Boyer-Lindquist (B-L) coordinates, we postulate to be of the form
\begin{multline}\label{4}
    \dd s^2 = \frac{G (F H+a^2 \cos ^2\theta ) \Psi }{(\sqrt{F} H+a^2 \sqrt{G} \cos ^2\theta )^2}\,
    \dd t^2-\frac{\Psi }{F H+a^2}\,\dd r^2+2 a \sin ^2\theta\Big[\frac{\sqrt{F} \sqrt{G} H-F G H}{(\sqrt{F} H+a^2 \sqrt{G} \cos ^2\theta )^2}\Big]\Psi\dd t\dd \phi\\
    -\Psi\dd \ta^2-\Psi\sin ^2\theta \Big\{1+a^2 \sin ^2\theta\Big[\frac{2 \sqrt{F} \sqrt{G} H-F G H
    +a^2 G \cos ^2\theta }{(\sqrt{F} H+a^2 \sqrt{G}\cos ^2\theta )^2}\Big]\Big\}\dd \phi^2,
\end{multline}
by solving the field equations for $\Psi(r,\ta)$, which depends also on the rotating parameter $a$. More on the derivation and generalization of~\eqref{4} will be given elsewhere~\cite{Azreg}. For fluids undergoing only a rotational motion about a fixed axis (the $z$ axis here), $T_{r\ta}\equiv 0$ leading to $G_{r\ta}= 0$, which is one of the very two equations to solve to obtain $\Psi(r,\ta)$. From now on, we use the following conventions and notations:  $\mu:\,1\leftrightarrow t,\,2\leftrightarrow r,\,3\leftrightarrow \ta,\,4\leftrightarrow \phi$) and $(u,\,e_2,\,e_3,\,e_4)=(u,\,e_r,\,e_{\ta},\,e_{\phi})$.

\section{The solutions \label{sec2}}

To ease the calculations, we use the algebraic coordinate $y=\cos\ta$ and replace $\dd \ta^2$ by $\dd y^2/(1-y^2)$ in~\eqref{4}. For the sake of subsequent applications (to regular \bh and wormholes), we will assume $H\neq r^2$ unless otherwise specified. Setting $K(r)\equiv \sqrt{F}H/\sqrt{G}$ and using an indexical notation for derivatives: $\Psi_{,ry^2}\equiv \partial^2\Psi/\partial r\partial y^2$, $K_{,r}\equiv \partial K/\partial r$, etc, the equation $G_{r\ta}= 0$ yields
\begin{equation}\label{5}
(K+a^2 y^2)^2 (3\Psi_{,r}\Psi_{,y^2} -2\Psi \Psi_{,ry^2}) =3a^2K_{,r}\,\Psi^2
\end{equation}
This hyperbolic partial differential equation may possess different solutions, but a simple class of solutions is manifestly of the form $\Psi(r,y)=g(K+a^2 y^2)$ where $g(z)$ is solution to
\begin{equation}\label{6}
    2z^2gg_{,zz}-3z^2{g_{,z}}^2+3g^2=0
\end{equation}
where $z=K(r)+a^2 y^2$. A general solution depending on two constants is derived setting $A(z)=g^{\,\prime}/g$ and leads to $\Psi_{\text{gen}}=c_2z/(z^2+c_1)^2$. However, this solution does not exhaust the set of all possible solutions of the form $g(z)$ to~\eqref{6} which, being nonlinear, admits other more interesting power-law solutions $g(z)\propto z^n$ leading to
\begin{equation}\label{8}
    \Psi_1=K(r)+a^2 y^2\;\text{ or }\;\Psi_2=[K(r)+a^2 y^2]^{-3}
\end{equation}
where $\Psi_2$ is included in $\Psi_{\text{gen}}$ taking $c_1=0$ and $c_2=1$. A consistency check of the field equations $G_{\mu\nu}=T_{\mu\nu}$ and the form of $T_{\mu\nu}$ [Eq.~\eqref{1}] yields the partial differential equation
\begin{equation}\label{8r}
\Psi[{K_{,r}}^2+K(2-K_{,rr})-a^2y^2(2+K_{,rr})]+(K+a^2y^2)(4y^2\Psi_{,y^2}-K_{,r}\Psi_{,r})=0,
\end{equation}
which is solved by $\Psi_1$ (but not by $\Psi_2$) provided $K=r^2+p^2$ where $p^2$ is real. We have thus found a simple common solution to both Eqs.~\eqref{5} and~\eqref{8r} given by
\begin{equation}\label{8ra}
    \Psi=r^2+p^2+a^2 y^2.
\end{equation}

We do not know the set of all possible solutions to Eqs.~\eqref{5} and~\eqref{8r}, however, we can still distinguish two families of rotating solutions. Depending on $G(r)$, $F(r)$ and $H(r)$, a rotating solution given by~\eqref{4} is called a normal fluid, $\Psi_n$, if the static solution~\eqref{3} is recovered from the rotating one in the limit $a\to 0$: This implies $\lim_{a\to 0}\Psi=H$. Otherwise the rotating solution is called a conformal fluid, $\Psi_c$. Given $G(r)$, $F(r)$ and $H(r)$, the normal $\dd s_n^2$ and conformal $\dd s_c^2$ fluids are conformally related
\begin{equation}\label{8a}
    \dd s_c^2=(\Psi_c/\Psi_n)\dd s_n^2.
\end{equation}
Now, since $\lim_{a\to 0}\Psi_c\neq H$ (by definition) and $\lim_{a\to 0}\dd s_n^2= \dd s_{\text{stat}}^2$ [Eq.~\eqref{3}], this implies that $\lim_{a\to 0}\dd s_c^2\neq \dd s_{\text{stat}}^2$, and thus $\lim_{a\to 0}\dd s_c^2$ is a new static metric conformal to $\dd s_{\text{stat}}^2$.

For the remaining part of this work, we shall explore the properties of both the normal (Sect.~\ref{sec3}) and conformal (Sect.~\ref{sec4}) rotating solutions that can be constructed using the unique simple solution $\Psi$ available to us, which is given by~\eqref{8ra}. From now on, we shall use the prime notation to denote derivatives of functions.

\section{Physical properties of the model-independent normal interior core: $G=F$ \label{sec3}}

The constraints $G=F$ and $K=r^2+p^2$ yield $H=K$, so we deal with a normal fluid since $\lim_{a\to 0}\Psi=H$ [Eq.~\eqref{8ra}]. The invariants $R$ and $R_{\mu\nu\al\bt}R^{\mu\nu\al\bt}$ are proportional to $\ro^{-6}$ and $\ro^{-12}$, respectively, with $\ro^2\equiv K+a^2y^2=H+a^2y^2$. Thus, the static and rotating solutions~\eqref{4} are regular if $H(r)$ is never zero ($p^2\neq 0$), which is the case for wormholes and some type of regular phantom \BH~\cite{phantom1,phantom2}. If $H=r^2$ ($p^2=0$), then the rotating solution~\eqref{4} may have a ring singularity in the plane $\ta=\pi/2$ ($y=0$) at $r=0$ (more details are given in~\cite{Azreg}). As we shall see below, there are cases where the numerators of $R$ and $R_{\mu\nu\al\bt}R^{\mu\nu\al\bt}$ also vanish for $r=0$ and $\ta=\pi/2$ to the same order, leading to a ring-singularity free solution~\eqref{4}. When this is the case, the components of the SET as well as the two invariants remain finite, but undefined, on the ring $\ro^2=0$.

Setting $2f(r)\equiv K-FH$, $\De(r)\equiv FH+a^2$ and $\Si\equiv (K+a^2)^2-a^2\De\sin ^2\theta$, the solution~\eqref{4} reduces to
\begin{align}
\label{8b}\dd s_n^2 = & \Big[1-\frac{2f}{\ro^2}\Big]\dd t^2-\frac{\ro^2}{\De}\,\dd r^2
    +\frac{4af \sin ^2\theta}{\ro^2}\,\dd t\dd \phi
    -\ro^2\dd \ta^2-\frac{\Si\sin ^2\theta}{\ro^2}\,\dd \phi^2\\
\label{8c}\quad = & \frac{\De}{\ro^2}\,(\dd t-a\sin^2\ta\dd \phi)^2-\frac{\sin^2\ta}{\ro^2}\,[a\dd t-(K+a^2)\dd \phi]^2-\frac{\ro^2}{\De}\,\dd r^2-\ro^2\dd \ta^2 .
\end{align}
We fix the basis $(u,\,e_r,\,e_{\ta},\,e_{\phi})$ by
\begin{equation}\label{10}
    u^{\mu}=\frac{(K+a^2,0,0,a)}{\sqrt{\ro^2 \De}}\,,\;
    e^{\mu}_r=\frac{\sqrt{\De}(0,1,0,0)}{\sqrt{\ro^2}}\,,\;
    e^{\mu}_{\ta}=\frac{(0,0,1,0)}{\sqrt{\ro^2}}\,,\;
    e^{\mu}_{\phi}=-\frac{(a\sin^2\ta,0,0,1)}{\sqrt{\ro^2}\sin\ta}.
\end{equation}
The components of the SET are expressed in terms of $G_{\mu\nu}$ as: $\ep=u^{\mu}u^{\nu}G_{\mu\nu}$, $p_r=-g^{rr}G_{rr}$, $p_{\ta}=-g^{\ta\ta}G_{\ta\ta}$, $p_{\phi}=e^{\mu}_{\phi}e^{\nu}_{\phi}G_{\mu\nu}$. We find:
\begin{align}
\label{11a}&\ep =\frac{2(rf^{\,\prime} -f)-p^2}{\rho ^4}+
\frac{2p^2(3 f-a^2 \sin ^2\theta)}{\rho ^6}\\
\label{11b}&p_r=-\ep -  \frac{2p^2\De}{\rho ^6}\,,\quad p_{\ta}=-p_r-\frac{f^{\,\prime\prime}}{\rho ^2}\,,\quad p_{\phi}=p_{\ta}+\frac{2p^2a^2\sin^2\ta}{\ro^6}.
\end{align}
Thus, for  wormholes and some type of regular phantom \BH~\cite{phantom1,phantom2} where always $\ro^2>0$ ($H$ never vanishes), the components of the SET are finite in the static and rotating cases. Eqs.~\eqref{11a} and~\eqref{11b} will be used in~\cite{Azreg} to derive the rotating counterpart of the stable exotic dust Ellis wormhole emerged in a source-free radial electric or magnetic field~\cite{ex}. If $H=r^2$, corresponding to regular as well as singular \BH, the above expressions reduce to those derived in~\cite{Gurses,Magli}: $\ep =-p_r=2(rf^{\,\prime}-f)/\ro^4$, $p_{\ta}=p_{\phi}=\ep -f^{\,\prime\prime}/\ro^2$. In this case the components of the SET diverge on the ring $\ro^2=0$ unless $f\propto r^4$ as $r\to 0$, resulting in $(1-F)\propto r^2$ as $r\to 0$, which corresponds to the (anti) de Sitter case and to regular \BH. In fact, most of regular \bh derived so far have de Sitter-like behavior near $r=0$~\cite{IrinaD,Lemos,regular}.

From the third Eq.~\eqref{11b}, on sees that the tangential pressures, ($p_{\ta},\,p_{\phi}$), are generally nonequal and are equal only if $p^2=0$ or/and if $a=0$ (the static case). Hence, in the general rotating case, the tensor $T^{\mu\nu}$ has four different eigenvalues representing thus a totally imperfect fluid.

It is straightforward to verify the validity of the continuity equation: $(\ep u^{\mu})_{;\mu}=0$, where the semicolon denotes covariant derivative. The conservation equation, $T^{\mu\nu}{}_{;\nu}=0$, is consistent with $u^{\mu}{}_{;\nu}u^{\nu}\neq 0$ which shows that the motion of the fluid elements is not geodesic. This is attributable to the nonvanishing of the $r$- and $\ta$-components of the pressure gradient.

The purpose of constructing rotating and nonrotating solutions with negative pressure components, as might be the case in~\eqref{11a} to~\eqref{11b}, is, as was made clear in~\cite{Magli}, two-fold, in that, following a suggestion by Sakharov and Gliner~\cite{Gliner1,Gliner2}, (1) the core of collapsing matter, with high matter density, should have a cosmological-type equation of state $\ep=-p$, (2) the problem of the ring singularity, which characterizes Kerr-type solutions, could be addressed if the interior of the hole is fitted with an imperfect fluid of the type derived above. Fitting the interior of the hole with a de Sitter fluid is one possible solution to the ring singularity~\cite{Magli,Lemos}. Another possibility is to consider a regular core or a conformal regular one as we shall see in the case $G\neq F$ (Sect.~\ref{sec4}).

\subsection{Rotating imperfect $\La$-fluid---de Sitter rotating solution}

Instances of application of~\eqref{4} to re-derive the Kerr-Newman solution from the \S solution and to generate a rotating imperfect $\La$-fluid (I$\La$F) from the \dS solution are straightforward. To derive the Kerr-Newman solution, we take $F=G=1-2m/r+q^2/r^2$ and $H=r^2$, the solution is then given by~\eqref{8b} with $2f_{\text{KN}}=2Mr-q^2$, $\De_{\text{KN}}= r^2+a^2-2Mr+q^2$, $\ro_{\text{KN}}^2=r^2+a^2\cos ^2\ta$ and $\Si_{\text{KN}}= (r^2+a^2)^2-a^2\De_{\text{KN}}\sin ^2\theta$.

Consider the de Sitter solution
\begin{equation}\label{f2a}
    \dd s_{\La}^2 = (1-\La r^2/3)\dd t^2 - (1-\La r^2/3)^{-1}\dd r^2 - r^2(\dd \ta^2+\sin^2\ta\dd \phi^2)
\end{equation}
where $F=G=1-\La r^2/3$ and $H=r^2$. The metric $\dd s_{\La}^2$ of the rotating I$\La$F is given by~\eqref{8b} with $2f_{\La}=\La r^4/3$, $\De_{\La}= r^2+a^2-\La r^4/3$, $\ro_{\La}^2=r^2+a^2\cos ^2\ta$ and $\Si_{\La}= (r^2+a^2)^2-a^2\De_{\La}\sin ^2\theta$. Except from a short description made in~\cite{Irina}, the rotating I$\La$F has never been discussed deeply in the scientific literature. The components of the SET are $\ep =\La r^4/\ro_{\La}^4$, $p_r=-\ep$, $p_{\ta}=p_{\phi}=-\La r^2(r^2+2a^2\cos ^2\theta)/\ro_{\La}^4$. The limit $a\to 0$ leads to de Sitter solution where the fluid is perfect with $\ep=\La$ and $p_r=p_{\ta}=p_{\phi}=-\La$.

The rotating I$\La$F is only manifestly singular on the ring $\ro_{\La}^2=0$ [$(\ta,\,r)=(\pi/2,\,0)$ or $(y,\,r)=(0,\,0)$]. In fact, the curvature and Kretchmann scalars
\begin{equation}\label{12}
R =-\frac{4 \Lambda  r^2 }{r^2+a^2 y^2}\,,\quad R_{\mu\nu\al\bt}R^{\mu\nu\al\bt}=\frac{8 \Lambda ^2 r^4 (r^8+4 a^2 y^2 r^6+11 a^4 y^4 r^4-2 a^6 y^6 r^2+6 a^8 y^8)}{3 (r^2+a^2 y^2)^6}
\end{equation}
do not diverge in the limit $(y,\,r)\to(0,\,0)$. Despite the fact that the limits do not exist, we can show that they do not diverge. Let $\mathcal{C}$: $r=a h(y)$ and $h(0)=0$ be a smooth path through the point $(y,\,r)=(0,\,0)$ in the $yr$ plane. We choose a path that reaches $(y,\,r)=(0,\,0)$ obliquely or horizontally but not vertically, that is, we assume that $h^{\prime}(0)$ is finite [for paths that may reach $(y,\,r)=(0,\,0)$ vertically we choose a smooth path $y=g(r)/a$ and $g(0)=0$ where $g^{\prime}(0)$ remains finite]. On $\mathcal{C}$, the limits of the two scalars as $y\to 0$ read
\begin{equation}\label{14}
    -\frac{4 \Lambda  h^{\prime}(0)^2}{1+h^{\prime}(0)^2}\,,\;\;\frac{8 \Lambda ^2 h^{\prime}(0)^4 [6
    -2 h^{\prime}(0)^2+11 h^{\prime}(0)^4+4 h^{\prime}(0)^6+h^{\prime}(0)^8]}{3 [1+h^{\prime}(0)^2]^6}\,,
\end{equation}
which are nonexisting [for $h^{\prime}(0)$ depends on the path] but they remain finite. Thus, the rotating I$\La$F is regular everywhere, however, the components of the SET are undefined on the ring $\ro^2=0$. Paths of the form: $y=g(r)/a$ and $g(0)=0$, where $g^{\prime}(0)$ remains finite, lead to the same conclusion. The other scalar, $R_{\mu\nu}R^{\mu\nu}$, behaves in the same way as the curvature and Kretchmann scalars.

Notice that the Kerr solution ($q=0$) and the rotating I$\La$F one are derived from each other on performing the substitution $2M\leftrightarrow \La r^3/3$, so that most of the Kerr solution properties, where no derivations with respect to $r$ are performed, are easily carried over into the rotating I$\La$F properties. For instance, the static limit, which is the 2-surface on which the timelike Killing vector $t^{\mu}=(1,0,0,0)$ becomes null, corresponds to $g_{tt}(r_{\text{st}},\ta)=0$ leading to $2\La r^2_{\text{st}}=3+\sqrt{9+12\La a^2\cos^2\ta}$. Thus, observers can remain static only for $r<r_{\text{st}}$. Similarly, the cosmological horizon, which sets a limit for stationary observers, corresponds to $\De_{\La}(r_{\text{ch}})=0$ leading to $2\La r^2_{\text{ch}}=3+\sqrt{9+12\La a^2}$.
Hence, the static limit is enclosed by the cosmological horizon and intersects it only at the poles $\ta=0$ or $\ta=\pi$ (in contrast with the Kerr solution where the static limit encloses the event horizon).

The four-velocity of the fluid elements may be expressed, in terms of the timelike $t^{\mu}$ and spacelike $\phi^{\mu}=(0,0,0,1)$ Killing vectors, as $u^{\mu}=N (t^{\mu}+\Om \phi^{\mu})$, with $N=(r^2+a^2)/\sqrt{\ro^2 \De_{\La}}$ and $\Om=a/(r^2+a^2)$ is the differentiable ($\Om\neq \text{constant}$) angular velocity of the fluid. Since the norm of the vector $t^{\mu}+\Om \phi^{\mu}$, $1/N^2$, is positive only for $\De_{\La}>0$, which corresponds to the region $r<r_{\text{ch}}$, the fluid elements follow timelike world lines only for $r<r_{\text{ch}}$. As $r\to r_{\text{ch}}$, $\Om$ approaches the limit $a/(r^2_{\text{ch}}+a^2)$ that is the lowest angular velocity of the fluid elements which we take as the angular velocity of the cosmological horizon: $\Om_{\text{ch}}=a/(r^2_{\text{ch}}+a^2)$. At the cosmological horizon, $t^{\mu}+\Om \phi^{\mu}$ becomes null and tangent to the horizon's null generators, so that the fluid elements are dragged with the angular velocity $\Om_{\text{ch}}$.

\section{Physical properties of the conformal interior core: $G\neq F$ \label{sec4}}

In this case $H\neq K=r^2+p^2$, unless $p^2=0$, leading to $\lim_{a\to 0}\Psi \neq H$. With $\Psi=K+a^2y^2$ [Eq.~\eqref{8ra}], the conformal rotating solution $\dd s_c^2$ is again given by~\eqref{8b} to~\eqref{8c} and the basis $(u,\,e_r,\,e_{\ta},\,e_{\phi})$ by~\eqref{10} but this time $\ro^2\equiv K+a^2y^2\neq H+a^2y^2$. The components of the SET are different due to the non-covariance of the field equations under conformal transformations~\cite{beyond}. The SET related to $\dd s_c^2$ is only partly proportional to that related to metric $\dd s_n^2$ and includes terms involving first and second order derivatives of the conformal factor $(K+a^2y^2)/(H+a^2y^2)$, which are the residual terms in the transformed Einstein tensor. Finally, the SET related to $\dd s_c^2$ takes the form
\begin{align}
\label{15a}&\ep =\frac{p^2 [6 f-r^2-p^2-a^2 (2-\cos^2\ta)]}{\rho ^6}+\frac{2 (r f^{\,\prime}-f)}{\rho ^4},
\quad p_{r}=-\ep -\frac{2 p^2 (r^2+p^2+a^2-2 f)}{\rho ^6}\\
\label{15c}&p_{\ta}=-\frac{2 (r^2+a^2 \cos^2\ta) f}{\rho ^6}+\frac{p^2+2 r f^{\,\prime}}{\rho ^4}-\frac{f^{\,\prime\prime}}{\rho ^2},\quad
p_{\phi}=p_{\ta}+\frac{2 a^2 p^2 \sin^2\ta}{\rho ^6}
\end{align}
which is finite and defined everywhere if $p^2\neq 0$. If $p^2=0$, the SET if finite, but undefined on the ring $\ro^{2}=0$, if $f\propto r^4$ as $r\to 0$ ((anti) de Sitter behavior for $F\neq G$). The curvature scalar
\begin{equation}\label{f1}
R=\frac{2 \{p^2 [r^2+p^2+a^2 (2-\cos^2\ta)]-2 p^2 f\}}{\rho ^6}-\frac{2 f^{\,\prime\prime}}{\rho ^2}
\end{equation}
is also finite for all $p^2$. The Kretchmann scalar is certainly finite everywhere for all $p^2$.
Conclusions made earlier concerning the continuity and conservation equations apply to the present case of the conformal fluid.

\subsection{Examples of static and rotating conformal imperfect fluids}

Consider a static regular \abh or a wormhole of the form~\eqref{3} where $G=F$ are finite at $r=0$ and $H(r)=r^2+q^2$. In the ($t,u,\ta,\phi$) coordinates, where $u$ is the new radial coordinate, $G(u)=G(r(u))$, $F(u)=G(u)/r^{\,\prime}(u)^2$ and $H(u)=r(u)^2+q^2$. Since we want $K(u)=u^2+p^2$ [Eq.~\eqref{8ra}], we have to solve the differential equation: $\dd r/\dd u=[r(u)^2+q^2]/(u^2+p^2)$, yielding
\begin{equation}\label{n1}
r(u)=q\tan[(q/p)\arctan(u/p)]
\end{equation}
where $p^2\neq 0$ and $q^2\neq 0$. In ($t,u,\ta,\phi$) coordinates, the equivalent static solution takes the form
\begin{equation}\label{f2b}
    \dd s_{(s)}^2 = G(u)\dd t^2 - \Big(\frac{r(u)^2+q^2}{u^2+p^2}\Big)^2\frac{\dd u^2}{G(u)} - [r(u)^2+q^2](\dd \ta^2+\sin^2\ta\dd \phi^2).
\end{equation}
While metrics~\eqref{3} and~\eqref{f2b} are equivalent, their rotating counterparts are not. The metric $\dd s_{c}^2$ of the conformal rotating core fluid, that is the rotating counterpart of~\eqref{f2b}, is given by~\eqref{8b} with $2f_{(s)}=u^2+p^2-F(u)H(u)$, $F(u)H(u)=(u^2+p^2)^2G(u)/[r(u)^2+q^2]$, $\De_{(s)}= F(u)H(u)+a^2$, $\ro_{(s)}^2=u^2+p^2+a^2\cos ^2\ta$ and $\Si_{(s)}= (r^2+p^2+a^2)^2-a^2\De_{(s)}\sin ^2\theta$. Since $p\neq 0$ the SET and curvature scalar, given by~\eqref{15a} to~\eqref{f1} on replacing $r$ by $u$, $f$ by $f_{(s)}$ and $\ro$ by $\ro_{(s)}$, are finite everywhere. One can thus follow one of the procedures in the literature~\cite{IrinaD,Magli,Lemos,regular}, as the one performed in~\cite{Magli}, to match the rotating metric $\dd s_{c}^2$ to the Kerr black hole.

It is straightforward to check that $\lim_{a\to 0}\dd s_{c}^2$ does not yield $\dd s_{(s)}^2$; rather, the limit yields a new static, conformal imperfect fluid, solution.

\section{Conclusion \label{sec5}}

A master metric in B-L coordinates that generates rotating solutions from static ones has been put forward. The final form of the generated stationary metric depends on a two-variable function that is a solution to two partial differential equation ensuring imperfect fluid form of the source term in the field equations. Only one simple solution of the two partial differential equations has been determined in this work and appears to lead to stationary, as well as static, normal and conformal imperfect fluid solutions.

On applying the approach to the de Sitter static metric and to a static regular \abh or a wormhole, two regular rotating, imperfect fluid cores, normal and conformal respectively, with equation of state nearing $\ep=-p$ in the vicinity of the origin ($r\to 0$), have been derived.

Conformal fluid cores have everywhere finite components of the SET and of the curvature and Kretchmann scalars.

We have not examined any energy conditions and related constraints on the mass density since even violations of the weak energy condition, not to mention the strong one, have become custom to issues pertaining to regular cores~\cite{Magli,cond1,cond2}. These violations worsen in the rotating case as was concluded in~\cite{Magli}.


\end{document}